\documentclass[twocolumn,pra,aps,showpacs]{revtex4}
\usepackage{amssymb}
\usepackage{amsmath}
\usepackage{epsfig}

\begin{document}

\title{Rosen-Zener Transition in a Nonlinear Two-Level System}
\author{ Di-Fa Ye$^{1,2}$}
\author{ Li-Bin Fu$^{1}$ }
\author{ Jie Liu$^{1,}$ }
\email[Email: ]{liu_jie@iapcm.ac.cn}

\affiliation{1.Institute of Applied Physics and Computational
Mathematics, Beijing
100088, P. R. China\\
2.Graduate School, China Academy of Engineering Physics, Beijing
100088, P. R. China}

\begin{abstract}
We study Rosen-Zener transition (RZT) in a nonlinear two-level system in
which the level energies depend on the occupation of the levels,
representing a mean-field type of interaction between the particles. We find
that the nonlinearity could affect the quantum transition dramatically. At
certain nonlinearity the $100\%$ population transfer between two levels is
observed and found to be robust over a very wide range of external
parameters. On the other hand, the quantum transition could be completely
blocked by a strong nonlinearity. In the sudden and adiabatic limits we have
derived analytical expressions for the transition probability. Numerical
explorations are made for a wide range of parameters of the general case.
Possible applications of our theory to Bose-Einstern Condensates (BECs) are
discussed.
\end{abstract}

\pacs{32.80.Qk,42.50.Vk,03.75.Lm} \maketitle

\section{Introduction}

The Rosen-Zener model was firstly proposed to study the spin-flip of
two-level (Hyperfine Zeeman energy level) atoms interacting with a rotating
magnetic field by N.Rosen and C.Zener to account for the double
Stern-Gerlach experiments \cite{RZT}. Different from the well-known
Landau-Zener model that depicts the tunnelling dynamics between two
avoided-crossing energy levels\cite{LZT}, in the Rosen-Zener model the
energy bias between two levels is fixed and the coupling between two modes
is time-dependent described by a rectangular \cite{Rabi}, Gaussian \cite%
{Gerald}, exponential \cite{Voronin} or a hyperbolic-secant functions\cite%
{RZT}. This model constantly attracts much attention not only because it has
exact analytic solution providing a bridge to understand complicated
multi-mode systems \cite{multimode} but also due to its versatile
applications in nonresonant charge exchange of ion-atom collision \cite%
{Olson}, laser-induced excitation \cite{excitation}, nuclear magnetic
resonance technique and quantum computation\cite{quntumcomp}, to name only a
few.

In the present paper, we extend the Rosen-Zener model to nonlinear
case and want to see how nonlinearity affects the quantum transition
dynamics in this system. Our work is motivated by recent flourish of
Bose Einstein condensates (BECs), where nonlinearity naturally
arises from a mean-field treatment of the interaction between
particles. In fact, nonlinear effects
constantly emerge in BEC system recently, such as self-trapping\cite%
{smerzi,wgf,self,kivshar}, superfluidity\cite{swallow}, instability\cite%
{instable} and nonlinear Landau-Zener tunneling\cite{wulzt,fixedpoint}.
Extending the famous Rosen-Zener model to nonlinear case is of great
interest and worthwhile at this point.

Our paper is organized as follows. In Sec.II, we introduce the
nonlinear Rosen-Zener model. In Sec.III, for the degenerate case
that the energy bias between two levels is zero, our numerical
calculations reveal the significant effects of nonlinearity on RZT.
We then derive analytic expressions for RZT in sudden and adiabatic
limits. In Sec.IV, our discussions are extended to nondegenerate
case. Numerical explorations for a wide range of parameters are
made. Interesting phenomena are presented and discussed. Finally, in
Sec.V, possible experimental realization of our model in
Bose-Einstern Condensates (BECs) are discussed.

\section{Nonlinear Rosen-Zener Model}

The nonlinear two-mode system we consider is described by following
dimensionless Schr\"{o}dinger equation,
\begin{equation}
i\frac{\partial }{\partial t}\binom{a}{b}=H(t)\binom{a}{b},  \label{seq}
\end{equation}%
with the Hamiltonian given by
\begin{equation}
{\small H(t)=[}\frac{\gamma }{2}+\frac{c}{2}(\left\vert b\right\vert
^{2}-\left\vert a\right\vert ^{2})]\hat{\sigma}_{z}+\frac{v}{2}\hat{\sigma}%
_{x},  \label{hamilton}
\end{equation}%
where $\hat{\sigma}_{x}$ and $\hat{\sigma}_{z}$ are Pauli matrices, $\gamma $
and $v$ are the energy bias and coupling strength between two modes,
respectively. $c$ is the nonlinear parameter describing the inter-atomic
interaction. The total probability $\left\vert a\right\vert ^{2}+\left\vert
b\right\vert ^{2}$ is conserved and set to be 1 without losing generality.

Different from the Landau-Zener type model where coupling keeps constant and
energy bias varies in time linearly\cite{LZT,wulzt,fixedpoint}, in this
model, the energy bias $\gamma $ and nonlinearity $c$ are set to be constant
whereas the coupling $v$ is time-dependent governed by an external pulse
field of the form \cite{comment},
\begin{equation}
v=\left\{
\begin{array}{l}
0,\text{ }t<0,t>T \\
v_{0}\sin ^{2}(\frac{\pi t}{T}),\text{ }t\in \lbrack 0,T]%
\end{array}%
\right. ,
\end{equation}%
where T is the scanning period of the external field.

In our following study, we assume the quantum state is prepared on
one mode initially. With the external field turned on, quantum
transition between different modes emerges. What we concern is the
population dynamics in the presence of the external field. The
transition probability $\Gamma $ is defined as the probability of
the particle occupying the other mode after the external field pulse
is over.

We start our analysis with the simplest case, i.e., both the energy
bias and nonlinear parameter vanish ($\gamma=0$ and $c=0 $). In this
case, the Schr\"odinger equation denoted by Eq.(\ref{seq}) is
solvable analytically. Setting $(a,b)=(1,0)$ as the initial
condition, we readily obtain the probability of the particle
populated on the other mode as the function of time,
\begin{equation}
p(t)\text{=}\left\vert b(t)\right\vert ^{2}\text{=}\sin ^{2}\text{(}v_{0}%
\frac{2\pi t-T\sin \frac{2\pi t}{T}}{8\pi }\text{)} ,  \label{probability}
\end{equation}
then total transition probability is obtained by substituting $t=T$ into the
above equation,
\begin{equation}
\Gamma =p(T)=\sin ^{2}(\frac{v_{0}T}{4}).  \label{transition}
\end{equation}

The above expression demonstrates a perfect Rabi-like oscillation of
the transition probability versus the pulse duration or scanning
period $T$ of the external field. The oscillation frequency is
proportional to the maximum coupling strength $v_0$.

\section{Nonlinear RZT for degenerate case ($\protect\gamma=0$)}

\subsection{General Properties}

With the emergence of nonlinearity, the transition dynamics dramatically
changes. In this case, the Sch\"odinger Eq.(\ref{seq}) is no longer
analytically solvable \cite{wuying}. We therefore exploit 4-5th Runge-Kutta
algorithm to trace the quantum evolution numerically and calculate the
transition probability meanwhile. In our calculation, we choose the maximum
coupling strength $v_{0}$ as the energy scale, thus the weak nonlinearity
and strong nonlinearity refer to $c/v_{0}<<1$ and $c/v_{0}>>1$, respectively.

\begin{figure}[t]
\begin{center}
\rotatebox{0}{\resizebox *{8.0cm}{9.0cm} {\includegraphics
{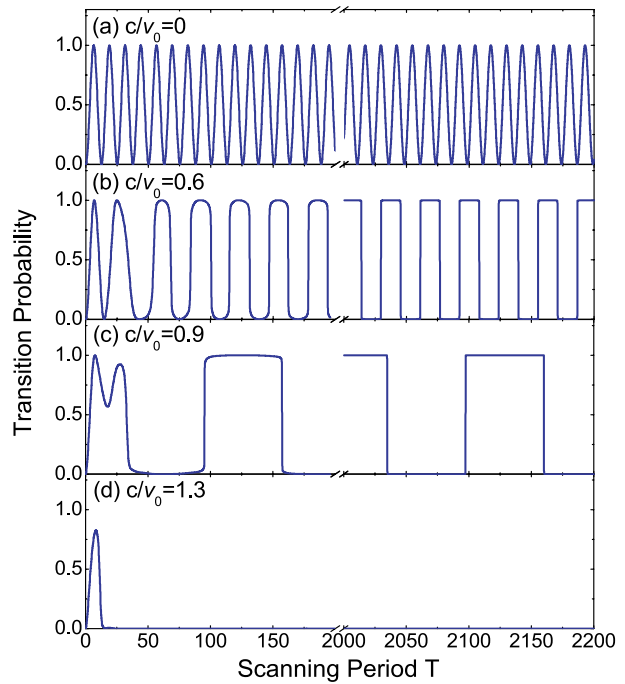}}}
\end{center}
\caption{(Color online.) Numerical results for the transition
probability versus the scanning period under different nonlinear
parameters $c/v_{0}$ (b) 0.6, (c) 0.9, (d) 1.3. For comparison, we
also include the result of linear case in (a), which well reproduces
the results predicted by formula (\ref{transition}).} \label{whole}
\end{figure}

Our numerical results are presented in Fig.\ref{whole}, where the
transition probability as the function of the scanning period are
plotted against the nonlinear parameters that range from weak
nonlinearity to strong nonlinearity. In Fig.\ref{whole}(a) we see a
regular oscillating pattern, agreeing with the analytic prediction
of formula (\ref{transition}). This regular periodic pattern is
destroyed with the emergence of nonlinearity. For weak nonlinear case $%
c/v_{0}<1$, the periodicity is lost only in short pulse regime, i.e., $%
0<T<50 $ at $c/v_{0}=0.9$. Whereas in the regime of large scanning
period, a rectangular periodic pattern revives instead. The period
of the rectangular oscillation increases with the enhancement of
nonlinearity. The above rectangular pattern is of particular
interest in practice because it represents a $100\%$ population
transfer between two modes robustly in a wide parameter regime.

For the case of strong nonlinearity, i.e., $c/v_{0}>1$ (see Fig.1d), the
quantum transition between two modes is even more affected by the
nonlinearity: The oscillation pattern is completely broken, and when the
scanning period $T > 15$ the quantum transition is totally blocked.

The above phenomena are interesting and intriguing, need detailed
investigation. Our further analysis includes two limit cases, namely,
adiabatic limit and sudden limit. The adiabatic limit means that the
external field varies slowly compared with the intrinsic motion of the
system. From the formula(\ref{transition}) we see the period of intrinsic
motion is characterized by $4\pi/v_{0}$ while the external field is
characterized by $T$. Thus, the adiabatic limit means $T>>4\pi/v_{0}$ or $%
v_{0}T>>4\pi$ whereas the sudden limit corresponds to
$v_{0}T<<4\pi$. In the following sections, we will further derive
some analytical formulas for the transition probability and explain
the above phenomena.

\subsection{Analytic Results for Sudden Limit Case($v_{0}T<<4\protect\pi$)}

\begin{figure}[t]
\begin{center}
\rotatebox{0}{\resizebox *{8.0cm}{4.0cm} {\includegraphics
{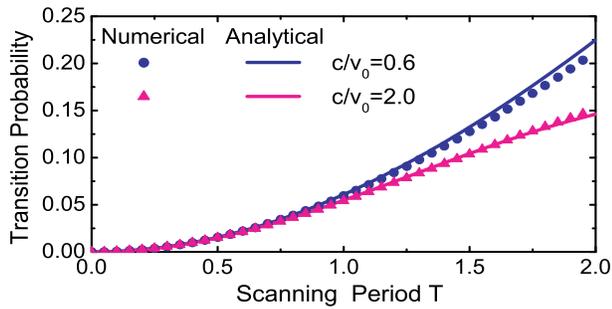}}}
\end{center}
\caption{(Color online). Transition probability in the sudden limit
for varied  nonlinear parameters. The scatters are obtained by
directly integrating Eq.(\protect\ref{seq}). The solid lines
demonstrate the analytical results based on
Eq.(\protect\ref{suddenformula}). They are in good agreement. }
\label{sudden}
\end{figure}

In the sudden limit that the scanning period of the external field
is small enough, it is clearly seen that nonlinearity always
suppress the transition from one mode to another mode (see
Fig.\ref{sudden}). This is because of the competition between the
on-site energy characterized by the nonlinear term and the hopping
energy characterized by the coupling strength. For a certain
coupling strength $v$, the larger is the nonlinear interaction $c$,
the larger is the on-site energy, therefore it will block the
population transfer between two modes more strongly and the
oscillation between two modes becomes slower.

Now we derive some analytic results supporting the above argument. Because
the transition probability is small, an explicit analytic expression can be
obtained using perturbation theory. We begin with the variable
transformation,
\begin{equation}
a=a^{^{\prime }}\exp [-i\int_{0}^{t}\frac{c}{2}(\left\vert b^{^{\prime
}}\right\vert ^{2}-\left\vert a^{^{\prime }}\right\vert ^{2})dt],
\label{transform1}
\end{equation}%
\begin{equation}
b=b^{^{\prime }}\exp [i\int_{0}^{t}\frac{c}{2}(\left\vert b^{^{\prime
}}\right\vert ^{2}-\left\vert a^{^{\prime }}\right\vert ^{2})dt].
\label{transform2}
\end{equation}%
As a result, the diagonal terms in the Hamiltonian are transformed away, and
we have
\begin{equation}
b^{^{\prime }}(t)=\int_{0}^{t}\frac{v_{0}}{2}\sin ^{2}\frac{\pi t}{T}%
e^{-i\int_{0}^{t}c(\left\vert b^{^{\prime }}\right\vert ^{2}-\left\vert
a^{^{\prime }}\right\vert ^{2})dt}a^{^{\prime }}dt.  \label{bpeqn}
\end{equation}%
The first-order amplitude of $b^{^{\prime }}(t)$ can be obtained by assuming
$a^{^{\prime }}=1,$ $b^{^{\prime }}=0$ on the right-hand side of the above
equation. This yields $b^{^{\prime }}(t)=\int_{0}^{t}\frac{v_{0}}{2}\sin ^{2}%
\frac{\pi t}{T}e^{ict}dt,$ and the transition probability equals to
\begin{equation}
\Gamma =\frac{2\pi ^{4}v_{0}^{2}[1-\cos (cT)]}{c^{2}(4\pi
^{2}-c^{2}T^{2})^{2}}.  \label{suddenformula}
\end{equation}%
For linear case of $c=0$, the above formula reduces to $\left\vert
b(T)\right\vert ^{2}=(\frac{v_{0}T}{4})^{2},$ which is the first order
expansion of the exact solution of transition probability of Eq.(\ref%
{transition}). For the nonlinear cases, this analytical
approximation is compared with our numerical results in
Fig.\ref{sudden} and good agreement is shown. Expanding
Eq.(\ref{suddenformula}) using $c$ as the small parameter gives,
\begin{equation}
\Gamma =(\frac{v_{0}T}{4})^{2}(1-\frac{c^{2}T^{2}}{12}),  \label{expand}
\end{equation}%
which indicates that the quantitative decrease of transition probability is
proportional to the square of nonlinear parameter $c$.

\subsection{Analytic Results for Adiabatic Limit Case($v_{0}T>>4\protect\pi$)%
}

\begin{figure}[t]
\begin{center}
\rotatebox{0}{\resizebox *{8.0cm}{3.0cm} {\includegraphics
{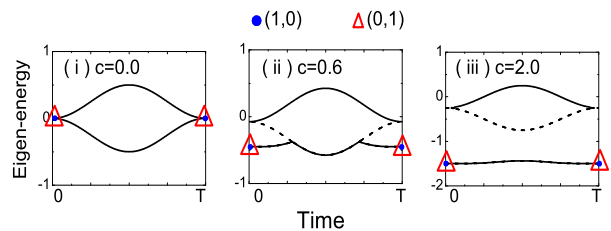}}}
\end{center}
\caption{Typical eigenenergy level structure in (a) linear case, (b)
weak nonlinear case, (c) strong nonlinear case. The dashed lines
correspond to unstable eigenstates \cite{pra}.} \label{eigenenergy}
\end{figure}

According to the adiabatic theory of nonlinear quantum mechanics
\cite{clasicalH}, the characters of quantum transition in the
adiabatic limit should be entirely determined by the  structure of
the energy levels and the properties of  corresponding eigenstates.
The eigenstates of the system satisfy that,
\begin{equation}
\lbrack \frac{c}{2}(\left\vert b\right\vert ^{2}-\left\vert a\right\vert
^{2})\hat{\sigma}_{z}+\frac{v}{2}\hat{\sigma}_{x}]\binom{a}{b}=\mu \binom{a}{%
b}.  \label{eigeneqn}
\end{equation}%
Solving the above nonlinear equations together with total particle
conservation condition $\left\vert a\right\vert ^{2}+\left\vert b\right\vert
^{2}=1$, we readily obtain the chemical potential $\mu $ and the eigenstate $%
(a,b)$. The eigenenergies can be derived according to the relationship $%
\epsilon =\mu -c/2 (\left\vert a\right\vert ^{4}+\left\vert
b\right\vert ^{4})$. Their dependence on the parameters is plotted
in Fig.\ref{eigenenergy} for  the cases of linearity, weak
nonlinearity and strong nonlinearity, respectively. Striking
phenomena are induced by the nonlinearity: Firstly, the  structure
of the energy levels change dramatically. In the linear case, there
are two  energy levels that are symmetric about a horizontal axis
(see Fig.3a). However, the symmetry breaks down in the presence of
nonlinearity and a new branch of eigenenergies emerges. For the weak
nonlinearity (i.e., Fig.3b),   at two ends and near the peak of the
field pulse there exists two levels, in other regime there are three
energy levels. When the nonlinearity is strong (i.e., Fig.3c), apart
from the two ends, the number of the energy levels are three.
Secondly, the eigenstates of the mid level( e.g. denoted  by the
dashed line in Fig.\ref{eigenenergy}) are unstable. This is
evaluated by investigating the Hamiltonian-Jaccobi matrix obtained
by linearizing the nonlinear equation (\ref{eigeneqn}) around the
eigenstates. The eigenvalues of the Hamiltonian-Jaccobi matrix can
be real, complex or pure imaginary. Only pure imaginary eigenvalues
correspond to stable states, others indicate the unstable
ones\cite{pra}.

The above changes of the energy level in the presence of
nonlinearity  is expected to affect the quantum transition between
levels greatly. However, because of the degeneracy of our concerned
states of  $(1,0)$ and $(0,1)$, from  the above diagram of the
energy levels we cannot draw a definite conclusion about  the
adiabatic evolution of the initial  state $(1,0)$. In the following,
we study an equivalent classical Hamiltonian instead, and with
analyzing its phase space we achieve insight into the adiabatic
evolution of the above nonlinear quantum system.
\begin{figure}[t]
\begin{center}
\rotatebox{0}{\resizebox *{8.0cm}{15.0cm} {\includegraphics
{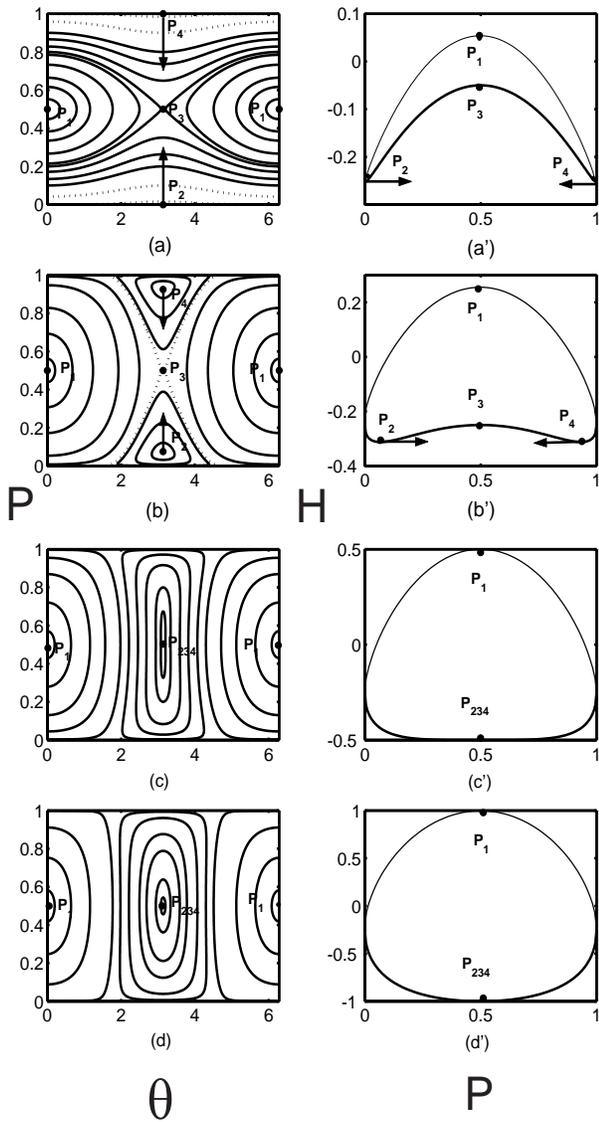}}}
\end{center}
\caption{Evolution of the phase space motions as $c/v$ changes
adiabatically (a)10, (b)2, (c)1, (d)0.5. The second column is the
corresponding energy curve at $\protect\theta=0$ (thin line) and
$\protect\pi$ (heavy line). The arrows indicate the shifting
direction of the fixed points as $v$ increases. } \label{phase}
\end{figure}

Following the theoretical methodology proposed in
Ref.\cite{clasicalH} we construct the effective classical
Hamiltonian with introducing two quantities: $p=\left\vert
b(t)\right\vert ^{2}$ as the probability of particles staying on
(0,1) mode , and $\theta =\theta _{a}-\theta _{b}$ as the relative
phase of the two modes. They form a pair of canonical variables of
following classical
Hamiltonian and satisfy the canonical equations, i.e., $%
d\theta/dt = \partial H/\partial s, ds/dt = -\partial H/\partial \theta$,
\begin{equation}
H=v\sqrt{p(1-p)}\cos \theta -\frac{c}{4}(2p-1)^{2}.  \label{classical}
\end{equation}
The above classical system is capable to totally describe the
dynamic properties of nonlinear quantum Rosen-Zener system (1) on a
projective Hilbert space\cite{clasicalH}. Its fixed points, i.e.,
energy extrema of the classical Hamiltonian, correspond to the
quantum eigenstates. For example, in Fig.\ref{phase},  the stable
elliptic fixed point $P_{1}$
 corresponds to the upper level of
Fig.\ref{eigenenergy}; the energy of $P_{2}$ and $P_{4}$ are
identical, therefore they correspond to the same energy level, i.e.,
the lower one in Fig.\ref{eigenenergy}; the saddle point $P_{3}$ is
unstable, corresponding to the mid level denoted by a dashed curve
in Fig.\ref{eigenenergy}. The adiabatic evolution of the quantum
eigenstates therefore can be evaluated by tracing the shift of the
classical fixed points in the phase space when the parameter $v$
varies in time slowly\cite{fixedpoint}.

The analytic expressions of the fixed points  are obtained from $\dot{p}=0$ and $%
\dot{\theta}=0$ :
\begin{equation}
\theta ^{\ast }=0,\pi ;
\end{equation}%
\begin{equation}
\frac{v(1-2p^{\ast })}{2\sqrt{p^{\ast }(1-p^{\ast })}}\cos \theta ^{\ast
}+c(1-2p^{\ast })=0.  \label{fixpoint}
\end{equation}%
The number of fixed points depends on the instantaneous coupling strength $v$
and the nonlinear parameter $c$. If $c/v<1$, there exists only two fixed
points $(\theta ^{\ast },p^{\ast })=(0,1/2),(\pi ,1/2)$. However, when $%
c/v>1 $, there exists four fixed points ($P_{1}$, $P_{2}$, $P_{3}$ and $%
P_{4} $ in Fig.\ref{phase}(a)(b) ) : $(\theta ^{\ast },p^{\ast
})=(0,1/2),(\pi ,1/2),(\pi ,\frac{1}{2}(1\pm \sqrt{1-\frac{v^{2}}{c^{2}}}))$%
. One of them ($P_{3}$) is a saddle point, while the other three ($P_{1}$, $%
P_{2}$ and $P_{4} $) are all elliptic points corresponding to the local
maximum ($P_{1}$) and minimum ($P_{2}$ and $P_{4}$) of the classical
Hamiltonian.

\begin{figure}[t]
\begin{center}
\rotatebox{0}{\resizebox *{8.0cm}{6.5cm} {\includegraphics
{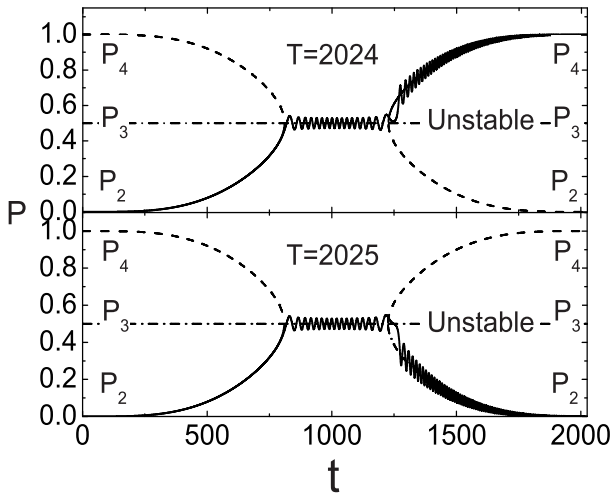}}}
\end{center}
\caption{Evolution of the particles and the fixed points at
$c/v_{0}=0.9$. Entirely different transition probabilities are
observed for a slight variation of the scanning period: (a) T=2024,
(b) T=2025. } \label{evl}
\end{figure}

When we  increase $v$, $P_{2}$, $P_{3}$ and $P_{4}$ will merge into
a new stable fixed point $P_{234}$ in a regime satisfying the
condition $v/c > 1$, as can be seen from
Fig.\ref{phase}(a$\rightarrow $b$\rightarrow $c$\rightarrow $d). An
interesting question arises when the parameter $v$ decreases to
below $c$ again: Which point will the  state choose to follow when
$P_{234}$ bifurcates
into $P_{2}$, $P_{3}$ and $P_{4}$ (Fig.\ref{phase}d$\rightarrow $c$%
\rightarrow $b$\rightarrow $a). The state that totally  follows
$P_{2}$ will  give a zero of the adiabatic transition probability,
whereas the state that totally follows the $P_{4}$ will correspond a
complete population transfer.  This classical picture could explain
why we see a rectangular pattern in Fig.\ref{whole}.

The above scenario is further supported by directly tracing the
evolution of particles as shown in Fig.\ref{evl}, where we also
demonstrate the temporal evolution of the fixed points. In the early
stage, we see that the state firmly follows the fixed point $P_{2}$.
It  starts to show a small  oscillation when  the fixed points
$P_{2}$, $P_{3}$ and $P_{4}$ merge. After that, the state either
follows the fixed point $P_{2}$ or $P_{4}$. The interesting thing is
that, at certain parameter, a slight change on  the period T could
thoroughly change the final transition probability, a signature of
the appearance of bistable state (Fig.\ref{phase}c$^{\prime
}\rightarrow $b$^{\prime }$).

Our study shows that whether  the state  follows $P_{2}$ or $P_{4}$
is determined by the total dynamical phase accumulated  during the
oscillation
motion around $P_{234}$, i.e. from $t^{\ast }$ to $t^{\ast \ast }$. Here, $%
t^{\ast }$ is the time when $P_{2}$, $P_{3}$ and $P_{4}$ merge into $P_{234}$%
, and  $t^{\ast \ast }$ is the moment when $P_{234}$ bifurcates into $P_{2}$%
, $P_{3}$ and $P_{4}$. They are obtained by setting $c=v(t^{\ast
})=v(t^{\ast \ast })$:

\begin{eqnarray}
t^{\ast } &=&\frac{T}{\pi }\sin ^{-1}\sqrt{\frac{c}{v_{0}}},  \notag \\
t^{\ast \ast } &=&T-t^{\ast }.  \label{criticaltime}
\end{eqnarray}

To obtain the total phase, we need to calculate the instantaneous
frequency that characterize the oscillations around the fixed point
at first. To this end, we expand the classical Hamiltonian around
the fixed point with $p=1/2+\delta _{p}$ , $\theta =\pi +\delta
_{\theta }$ , leading to

\begin{equation}
\delta H=\frac{1}{4}(v-c)\delta _{p}^{2}+\frac{1}{4}v\delta _{\theta }^{2},
\label{deltaH}
\end{equation}
with ignoring the higher order terms.  The instantaneous frequency
is then derived as
\begin{equation}
\omega (t)=\frac{1}{2}\sqrt{v(t)[v(t)-c]}.  \label{frequency}
\end{equation}
Integrating $\omega (t)$ from $t^{\ast }$ to $t^{\ast \ast }$ gives the
total phase
\begin{equation}
\varphi =\int_{t^{\ast }}^{t^{\ast \ast }}\omega (t)dt=\frac{v_{0}-c}{4}T.
\label{totalphase}
\end{equation}%
This expression indicates that the total phase increases linearly
with the scanning period. A $\pi$-value change on the phase will
change the choice of the state that either follows fixed point
$P_{2}$ or $P_{4}$.  Thus the period of rectangular oscillation
observed in Fig.\ref{whole} under adiabatic limit can be expressed
as
\begin{equation}
T_{rectangular}=4\pi /(v_{0}-c).  \label{rectangularperiod}
\end{equation}%

\begin{figure}[t]
\begin{center}
\rotatebox{0}{\resizebox *{8.0cm}{4.5cm} {\includegraphics
{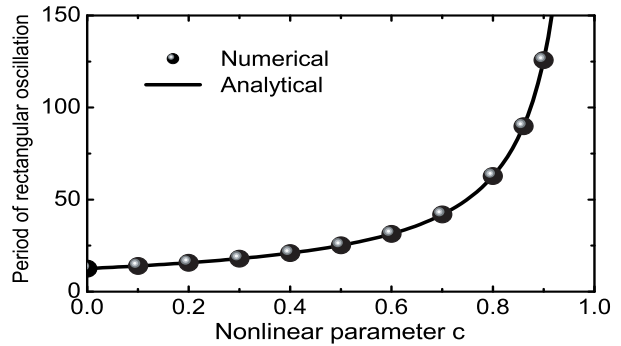}}}
\end{center}
\caption{Periods of rectangular oscillation as shown in Fig.\protect\ref%
{whole} under adiabatic limit. Numerical results are obtained by
directly observing Fig.\protect\ref{whole} while the analytical
curve is the plot of function $4\protect\pi/(v_{0}-c)$, here
$v_{0}=1$. They agree well with each other.} \label{period}
\end{figure}

To check the above theory, we have numerically solve  the nonlinear
Schr\"odinger equation for a wide range of the parameters. The
comparison between the analytical result  and the  numerical data
shows a good agreement in Fig.\ref{period}.

The complete suppression of quantum transition under adiabatic limit in Fig.\ref%
{whole}(d) can also be explained from the above picture. We briefly
state it as follows. For the strong nonlinearity that $c/v_{0}>1$,
the phase space evolution only undergoes
Fig.\ref{phase}(a$\rightarrow $b$\rightarrow $a) as $v$ increases
and decreases. During the process, no collision between the fixed
points occurs. Thus the state initially populated on $P_{2}$ can
safely remain on the fixed point, and finally come back to the
original  state smoothly. Thus no transition is observed.

\begin{figure}[t]
\begin{center}
\rotatebox{0}{\resizebox *{8.0cm}{16cm} {\includegraphics
{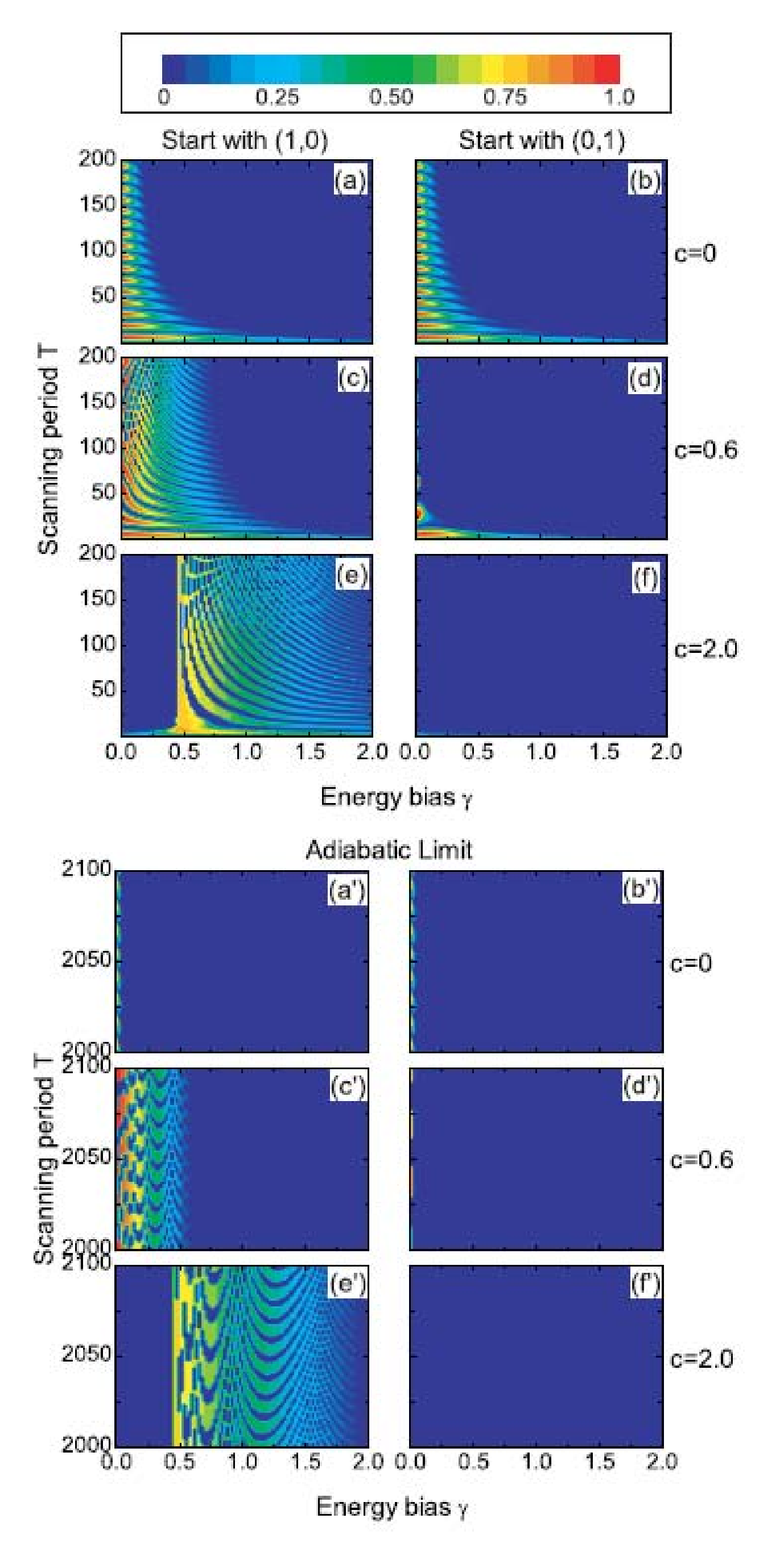}}}
\end{center}
\caption{(color online.) The contour plot of RZT probability as the function
of energy bias and scanning period at different nonlinear parameters (upper
panel) and its adiabatic counterpart (lower panel). Initially, all the
particles are prepared on the high energy mode (1, 0) (left column) or low
energy mode (0, 1) (right column), respectively, before the external field
is turned on. }
\label{gamaall}
\end{figure}

\section{Nonlinear RZT for Nondegenerate Cases ($\protect\gamma\neq 0$)}

In this section, we extend our discussions to the nondegenerate case ($%
\gamma\neq 0$). The transition probability is contour plotted with
respect to the energy bias and scanning period for different
nonlinear parameters ranging from weak nonlinearity to strong
nonlinearity in Fig.\ref{gamaall}.

The plotting reveals main features of the RZT in the nondegenerate
case: i) From the first panel of Fig.\ref{gamaall}, the contour plot
shows fringe structure: The bright zones correspond to high
transition probability whereas dark areas indicate low transition
probability. The fringe pattern is formed because of the periodic
behavior of the RZT with respect to the scanning period. For linear
(i.e., Fig.\ref{gamaall}a) and weak nonlinear cases (i.e.,
Fig.\ref{gamaall}c), this periodicity is broken by a larger energy
bias; For strong nonlinear case (i.e., Fig.\ref{gamaall}e), the
periodicity is absent both for larger and smaller energy bias.
ii)The nonzero energy bias leads to the asymmetry of RZT plottings
for the initial states of  $(1,0)$  and $(0,1)$. This is clearly
seen by comparing the right column with the left column of Fig.6.
iii) The second panel of Fig.6 shows the contour plot of transition
probability in the adiabatic limit i.e., very large scanning period
$T$. In this situation, the fringe structure is confined in a zonary
region. This zonary region is of particular interest in practice
because only in this region a robust population transfer between two
modes is possible. Outside the zonary region, the transition
probability almost equals to zero. Interestingly, the area of the
zonary region broaden
with increasing the nonlinear parameter (Fig.\ref{gamaall}a$%
^{\prime}\rightarrow$c$^{\prime}\rightarrow$e$^{\prime}$). When the
nonlinearity is strong (i.e., Fig.6e'), the zonary region shift away
from the origin. That is to say, to realize robust population
transfer for the highly nonlinear case, it requires a large energy
bias between two modes to compensate the difference of the on-site
energy caused by the nonlinearity.
iv) In the zonary region, for a
fixed $\gamma$, the transition probability oscillates rectangularly
versus the scanning period similar to the behavior shown in Fig.1.
But the amplitude of the oscillation decreases monotonously with
increasing the energy bias $\gamma$. This is indicated by the
gradual darkness from left to right side in the figures. The
correspondence between oscillating amplitude and energy bias
suggests that the probability of the population transferred to the
other mode can be designed at will by tuning the energy bias.

\begin{figure}[t]
\begin{center}
\rotatebox{0}{\resizebox *{8.0cm}{3.0cm} {\includegraphics
{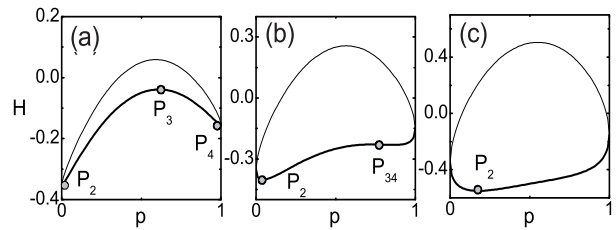}}}
\end{center}
\caption{Energy curve of $H$ ( Eq.(\protect\ref{classical}) ) at $c=1,
\protect\gamma=0.1$, $\protect\theta=0$ (thin line) or $\protect\pi$ (heavy
line), $v=$ (a)0.1, (b)0.5, (c)1.0. With the increase of $v$, $P_{3}$ and $%
P_{4}$ collide with each other and disappear subsequently, while $P_{2}$
maintains during the whole process.}
\label{energygama}
\end{figure}

We can achieve insight into the above findings by analyzing the evolution of
the fixed points corresponding to energy extrema. The energy curve in Fig.%
\ref{energygama} is tilted by the energy bias (for comparison one can recall
Fig.\ref{phase}), directly leading to the asymmetry on the diagram of
transition probability. The reason is stated as follows. Based on our
pictures in Sec.III(C), in the adiabatic limit, the transition between two
modes is always accompanied by a collision between fixed points. However,
with increasing $v$, the fixed point corresponding to the low energy mode ($%
P_{2}$) do not collide with any other fixed points, thus there is no
transition from low energy mode to higher one in the adiabatic
limit. The situation is quite different for the high energy mode
($P_{4}$), which collides with the unstable state ($P_{3}$) and
disappear subsequently as shown in Fig.\ref{energygama}(b). The
condition for the occurrence of the collision is given by
\cite{fixedpoint},
\begin{equation}
v=(c^{\frac{2}{3}}-\gamma ^{\frac{2}{3}})^{\frac{3}{2}}  \label{criticalgama}
\end{equation}
Considering $v$ can only varies in the interval $[0,v_{0}]$, we therefore
determine analytically the boundaries of the zonary regions as, $[0,c]$ if $%
c<v_{0}$ and $[(c^{\frac{2}{3}}-v_{0} ^{\frac{2}{3}})^{\frac{3}{2}},c]$ if $%
c>v_{0}$. These two analysis is in a good agreement with the
numerical results ( see Fig.\ref{energygama}(c$^{\prime
}$)(e$^{\prime }$) ).

Our above discussions assume that the nonlinear parameters are positive that
corresponds to the repulsive interaction between particles. For the
attractive interaction, the transition from (1,0) mode to (0,1) mode is
equivalent to that transition from (0,1) mode to (1,0) mode for the
repulsive interaction. Therefore, our discussions can be readily extended to
the negative-nonlinearity case that corresponds to attractive interaction
between particles. The detailed discussion is not repeated.

\section{Applications and Discussions}

Our model is applicable to describe the quantum transition of
two-mode BECs. One example is the BECs in a double-well. In such
system, the wavefunction can be described by a superposition of two
states that localize in each well separately. Then, the transition
dynamics of BECs between the two wells is well described by
Eq.(1)\cite{self}. The optical double-well can be created, for
example, by superimposing a blue-detuned laser beam upon the center
of the magnetic trap \cite{2well}. In this case, $\gamma$ denotes
the difference of the zero-point energy between two wells, $c $ is
the interaction between BEC atoms that can be adjusted flexibly by
Feshbach resonance, $v$ represents the height of the barrier that
separate the two wells. The barrier height can be effectively
controlled by adjusting the intensity of the blue-detuned laser
beam. Initially, we upload BECs atoms into one well, then ramp up
and down the barrier slowly, the nonlinear Rosen-Zener transition
should be observed.

Another promising candidate to observe nonlinear Rosen-Zener
transition is a spinor BEC in an optical trap. In such case, a
near-resonant field is applied
to the condensates to couple two hyperfine states of $^{87}$Rb, e.g., $%
\left\vert F=1,m_{F}=-1\right\rangle $ and $\left\vert
F=2,m_{F}=+1\right\rangle $ like in Ref.\cite{spinor}. Within the
standard rotating wave approximation, the system Hamiltonian can be
cast into the similar form of Eq.(\ref{hamilton}) \cite{Leggett},
where $\gamma $ denotes the frequency detuning, $c$ comes from the
contributions of both the homo-species and inter-species $s$-wave
scattering, $v$ is proportional to the intensity of the
near-resonant laser field. After preparing the BEC atoms in one
internal state, a sin$^{2}$ enveloped laser pulse is shined on the
condensates, the nonlinear Rosen-Zener transition is expected to
emerge.

In conclusion, we have investigated the Rosen-Zener transition in a
nonlinear two-level system and show that the nonlinearity could dramatically
affect the transition dynamics leading to many interesting phenomena. The
nonlinear RZT is suggested to be observed in the two-mode BEC systems.

\bigskip

\section{Acknowledgments}

This work is supported by National Natural Science Foundation of China
(No.10725521,10604009), the National Fundamental Research Programme of China
under Grant No. 2006CB921400, 2007CB814800.


\begin{thebibliography}{*}
\bibitem{comment} The sin$^{2}$ enveloped field is usually used to simulate
laser pulse. We use it to demonstrate our theory, however our main results
do not sensitively depend on the shape of the pulse.
\end{thebibliography}
\end{document}